\documentclass[journal]{IEEEtran}
\usepackage{bm}
\usepackage{amssymb}
\usepackage{amsfonts}
\usepackage{amsmath}
\usepackage{algorithm}
\usepackage{balance}
\usepackage{graphicx,subfigure,cite,multicol,multirow,diagbox,booktabs,array}
\usepackage{epstopdf}

\usepackage{color}
\usepackage{float}
\usepackage{stfloats}
\usepackage{makecell}
\ifCLASSINFOpdf
\else
\fi
\begin{document}
\title{Low-Complexity Three-Dimensional AOA-Cross Geometric Center Localization Methods via Multi-UAV network}

\author{Baihua Shi, Yifan Li, Guilu Wu, Shihao Yan, and Feng Shu
	
\thanks{This work was supported in part by the National Natural Science Foundation of China (Nos.U22A2002, and 62071234), the Major Science and Technology plan of Hainan Province under Grant ZDKJ2021022, and the Scientific Research Fund Project of Hainan University under Grant KYQD(ZR)-21008.}
\thanks{B. Shi and Y. Li are with the School of Electronic and Optical Engineering, Nanjing University of Science and Technology, Nanjing 210094, China. (e-mail: shibh56h@foxmail.com)}
\thanks{F. Shu is with the School of Information and Communication Engineering, Hainan University, Haikou, 570228, China and also with the School of Electronic and Optical Engineering, Nanjing University of Science and Technology, Nanjing, 210094, China. (e-mail: shufeng0101@163.com)}
\thanks{G. Wu is with the School of Information and Communication Engineering, Hainan University, Haikou, 570228, China.}
\thanks{S. Yan is with the School of Science and Security Research Institute, Edith Cowan University, Perth, WA 6027, Australia(e-mail: s.yan@ecu.edu.au).}
}
\maketitle

\begin{abstract}
Angle of arrival (AOA) is widely used to locate a wireless signal emitter in unmanned aerial vehicle (UAV) localization. Compared with received signal strength (RSS) and time of arrival (TOA), it has higher accuracy and is not sensitive to time synchronization of the distributed sensors. However, there are few works focused on three-dimensional (3-D) scenario. Furthermore, although maximum likelihood estimator (MLE) has a relatively high performance, its computational complexity is ultra high. It is hard to employ it in practical applications. This paper proposed two multiplane geometric center based methods for 3-D AOA in UAV positioning. The first method could estimate the source position and angle measurement noise at the same time by seeking a center of the inscribed sphere, called CIS. Firstly, every sensor could measure two angles, azimuth angle and elevation angle. Based on that, two planes are constructed. Then, the estimated values of source position and angle noise are achieved by seeking the center and radius of the corresponding inscribed sphere. Deleting the estimation of the radius, the second algorithm, called MSD-LS, is born. It is not able to estimate angle noise but has lower computational complexity. Theoretical analysis and simulation results show that proposed methods could approach the Cramer-Rao lower bound (CRLB) and have lower complexity than MLE.
\end{abstract}


\IEEEpeerreviewmaketitle

\section{Introduction}
Source location has attracted much attention for decades. Especially, wireless sensor network (WSN) and integrated sensing, internet of things (IoT), unmanned aerial vehicle (UAV) localization and communications (ISAC) are emerging in recent years \cite{shuDMRIS2021tcom,wang2015twc,Li2021CL,LiuISAC2020tcom,shuDMRIS2021tcom,shiRIS2021cl,dongRIS2022jcn}. As massive multiple-input multiple-output (MIMO) became widespread, it is easy to achieve angles information. Thus, angle of arrival (AOA) is becoming popular in recent years.

In the literature, there are two kinds of passive wireless signal positioning algorithms \cite{Wufp2015tvt}. 
The first kind is called scene analysis. It needs to know channel state information (CSI) of all cells in advance. Then, locate the source by matching the received CIS with CSI in database. Thus, there is an offline training phase in this system \cite{Wufp2015tvt}. That means its accuracy is related to training.
However, it is low-cost and could be well used in the non-line-of-sight (NLOS) condition. Thus, it is mainly employed in indoor positioning, like fingerprint-based methods \cite{cheng2022wcl}.

The second kind of localization methods is triangulation. Time of arrival (TOA), the time difference of arrival (TDOA) \cite{HoTDOA2012tsp,Shu2018SJ}, received signal strength (RSS) \cite{Jin2020TSP,Li2021CL} and AOA \cite{Lingren1978tae,wang2015twc,wang2018twc} all fall into that. This kind of systems could be used in both indoor and outdoor. Among them, TDOA is a modified version of TOA. The famous global positioning system (GPS) locate source by adopting TOA. In \cite{Bishop2009tae}, authors focused on 2-D positioning. Maximum likelihood estimator (MLE) for RSS, TOA and AOA was derived. Furthermore, estimator accuracy was also presented. Two bias reduction methods were proposed in \cite{HoTDOA2012tsp} to improve the accuracy. To provide a guidance, performance of TDOA in multi-satellite localization was analyzed in \cite{Shu2018SJ}. 
Since the sensor could be equipped with only one antenna, RSS is the lowest cost scheme. Thus, many RSS-based algorithms have been studied for practical applications, such as cellular location \cite{WeissRSS2003tvt}, UAV localization \cite{Li2021CL} and so on.

Different from that TOA require accurate time synchronization between sensors and RSS is sensitive to the difference of antenna pattern in all sensors, AOA just needs sensors to be equipped with multi antennas, which is not a threshold as development of MIMO. In practical applications, UAV network will detect target firstly \cite{jieDOA2022wcl}. Then, angles could be measured by Direction-ofArrival (DoA) estimation \cite{shuDOA2018tcom,shiHADDOA2022scis,shiDOA2022sj,shiDOA2022wcl,chenDOA2022wcl}. There are many closed-form estimators for 2-D AOA \cite{Lingren1978tae,wangAOA2012tsp,wang2DAOA2013wcl,lu2DAOA2012twc} and a few methods for 3-D scene \cite{kutluyAOA2015tsp,wang2015twc,wang2018twc,guiAOA2023iot}. However, 2-D approaches can not be applied in 3-D scene. And, the bias of sensors is not considered. Thus, authors proposed a bias-free closed-form algorithm for 3-D AOA in \cite{wang2015twc}. Moreover, the corresponding Cram\'{e}r-Rao lower bound (CRLB) was derived to provide a reference. In \cite{wang2018twc}, 
authors pointed out that Hybrid Bhattacharyya-Barankin (HBB) Bound is more suitable than CRLB to be a benchmark when noise level is high. A hybrid AOA-TDOA positing estimator was also proposed. In addition, when the source is far enough, the scene is far-field and AOA positioning will convert to a DOA estimation. In \cite{guiAOA2023iot}, a noval 3-D AOA method for UAV positioning was proposed by adopting bistatic MIMO radar. Transmitting and receiving array measure the 2D angle-of-departure and angle-of-arrival respectively. Then, the 3D position of the UAV is calculated by these angles.

However, existed high performance estimators for 3-D AOA localization have relatively high computational complexity. The proposed method in \cite{wang2015twc} needs to have a prior knowledge of noise variance, which is unavailable in practical applications.
Thus, in this paper, we aim to develop high performance and low complexity 3-D AOA methods for 3-D AOA positioning.
The main contributions of this paper are summarized as follows:
\begin{enumerate}
	\item 3-D AOA positioning is considered in this paper. Inspired by the inscribed sphere of a tetrahedron, we extend every angle to a plane. Then, the 3-D AOA localization can be transferred into the seeking for a center of inscribed sphere for these multi planes. Our method not only could accurately estimate the source position and the noise level at the same time but also has very low computational complexity, which is similar to the conventional least square (LS) estimator.
	\item To reduce more computational complexity, the estimation of noise level is removed. Then, the second algorithm is born. The original optimization problem could be converted to a LS problem. The simulation results show that these two methods have similar performance. Furthermore, compared with conventional LS, proposed methods have about 8 dB gain.
	\item The CRLB and computational complexity are presented. Theoretical analysis and simulation results respectively unveiled that the performance of proposed methods is close to CRLB and the computational complexity is reduced significantly. The propsoed methods could achieve a satisfactory balance between the performance and computational complexity. 
\end{enumerate}
\emph{Notations:} Throughout the paper, vectors and matrices are respectively denoted by $\mathbf{x}$ and $\mathbf{X}$ in bold typeface, while normal typeface is used to represent scalars, such as $x$. Signs $(\cdot)^T$, $(\cdot)^H$, $|\cdot|$ and $\|\cdot\|$ represent transpose, conjugate transpose, modulus and norm, respectively. $\mathbf{I}_M$ denotes the $M\times M$ identity matrix. Furthermore, $\mathbb{E}[\cdot]$ represents the expectation operator, and $\mathbf{x}\sim \mathcal{CN}(\mathbf{m},\mathbf{R})$ denotes a circularly symmetric complex Gaussian stochastic vector with mean vector $\mathbf{m}$ and covariance matrix $\mathbf{R}$.
$\hat{x}$ denotes the estimated value of $x$.

The rest of this paper is organized as follows. The system model of the 3D-AOA positioning for Multi-UAV network is presented in Section II. Then, two high performance and low complexity methods are proposed in Section III. The performance accuracy and computational complexity are investigated in Section IV. In Section V, simulation and numerical results are provided to analyze the performance and convergence of proposed methods. Finally, we come to the conclusion in Section VI.
\section{System Model}
\begin{figure}
	\centering
	\includegraphics[width=0.48\textwidth]{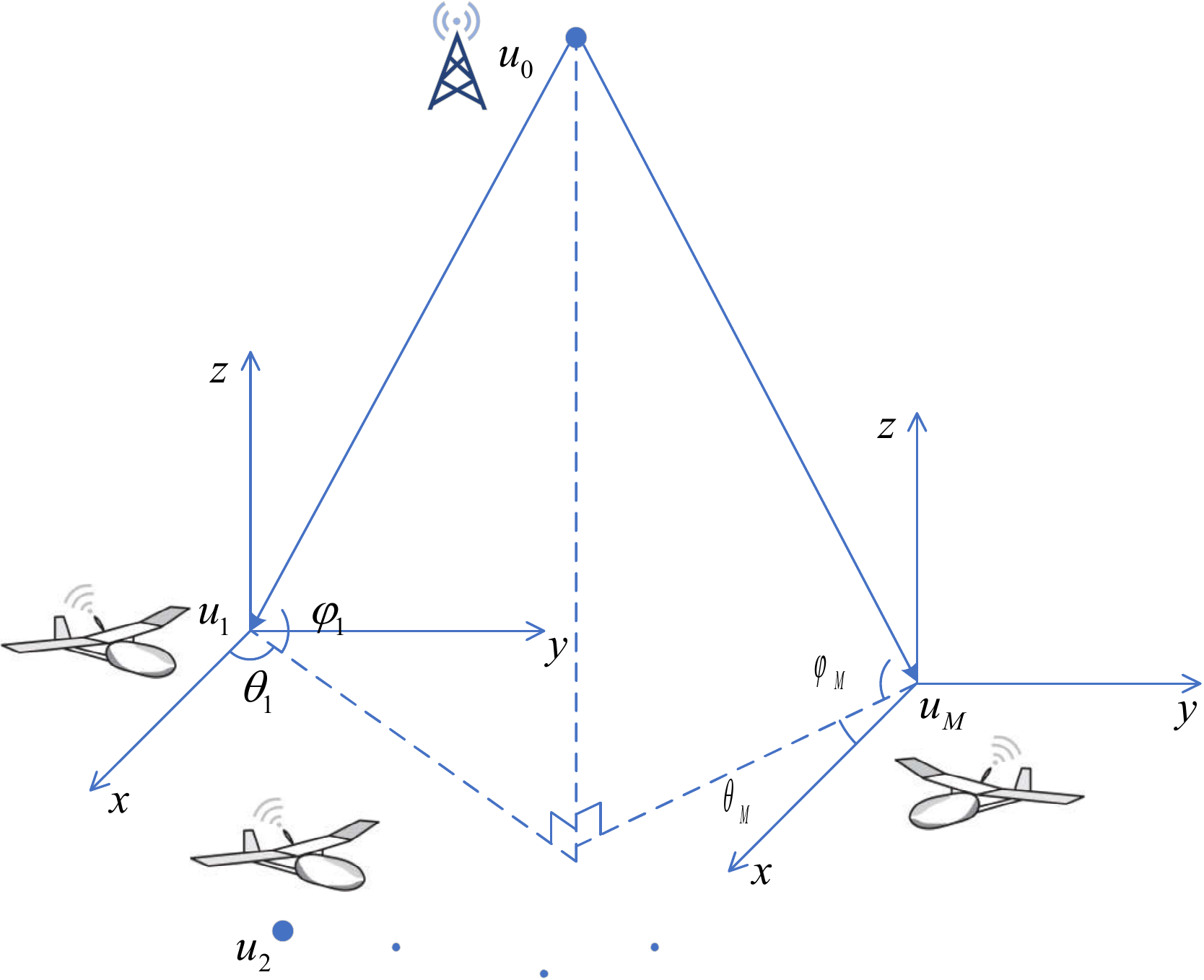}\\
	\caption{System model for 3-D AOA localization}
	\label{fig_sys}
\end{figure}
In this paper, a 3-D AOA localization system with $M$ UAVs is considered. Every UAV is able to measure the azimuth angle and elevation angle of the signal transmitted by the source.

As shown in Figure~\ref{fig_sys}, the position of the source and UAVs is defined by $\mathbf{u}_0=[x_0,y_0,z_0]^T$ and
$\mathbf{u}_m=[x_m,y_m,z_m]^T,~m=1,2,\cdots,M$, respectively.
For $m$th UAV, the azimuth angle $\theta_m$ and elevation angle $\phi_m$ related to the coordinate of UAVs and source can be expressed by
\begin{align}\label{angle1}
	\theta_m=\arctan\left(\frac{y_m-y_0}{x_m-x_0}\right),
\end{align}
\begin{align}\label{angle2}
	\phi_m=\arctan\left(\frac{z_m-z_0}{(x_m-x_0)\cos\theta_m+(y_m-y_0)\sin\theta_m}\right)
\end{align}
where $\theta_m\in(-\pi,\pi)$ and $\theta_m\in(-\pi/2,\pi/2)$.

In practical, the angles estimated by UAVs contain noise. Thus, we denote azimuth angles $\bm{\theta}_r$ and elevation angles $\bm{\phi}_r$ as
\begin{align}\label{Angle}
	\bm{\theta}_r=\boldsymbol{\theta}+\mathbf{n}_{\boldsymbol{\theta}}
\end{align}
\begin{align}\label{Angle2}
	\bm{\phi}_r=\boldsymbol{\phi}+\mathbf{n}_{\boldsymbol{\phi}}
\end{align}
where $\bm{\theta}=\left[\theta_1, \theta_2,\cdots, \theta_M\right]^T$ and $\bm{\phi}=\left[\phi_1,\phi_2,\cdots \phi_M\right]^T$ are true values. $\mathbf{n}_{\boldsymbol{\theta}} = [n_{\theta_1},n_{\theta_2},\cdots,n_{\theta_M}]^T$ and $\mathbf{n}_{\boldsymbol{\phi}}= [n_{\phi},n_{\phi},\cdots,n_{\phi}]^T$ are the additive zero mean Gaussian noises, where $\mathbf{n}_{\boldsymbol{\theta}}\sim \mathcal{CN}(0,\sigma^2_{\theta}\mathbf{I}_M)$ and $\mathbf{n}_{\boldsymbol{\phi}}\sim \mathcal{CN}(0,\sigma^2_{\phi}\mathbf{I}_M)$.

\section{Geometric Center based methods for AOA localization}
\begin{figure}
	\centering
	\includegraphics[width=0.37\textwidth]{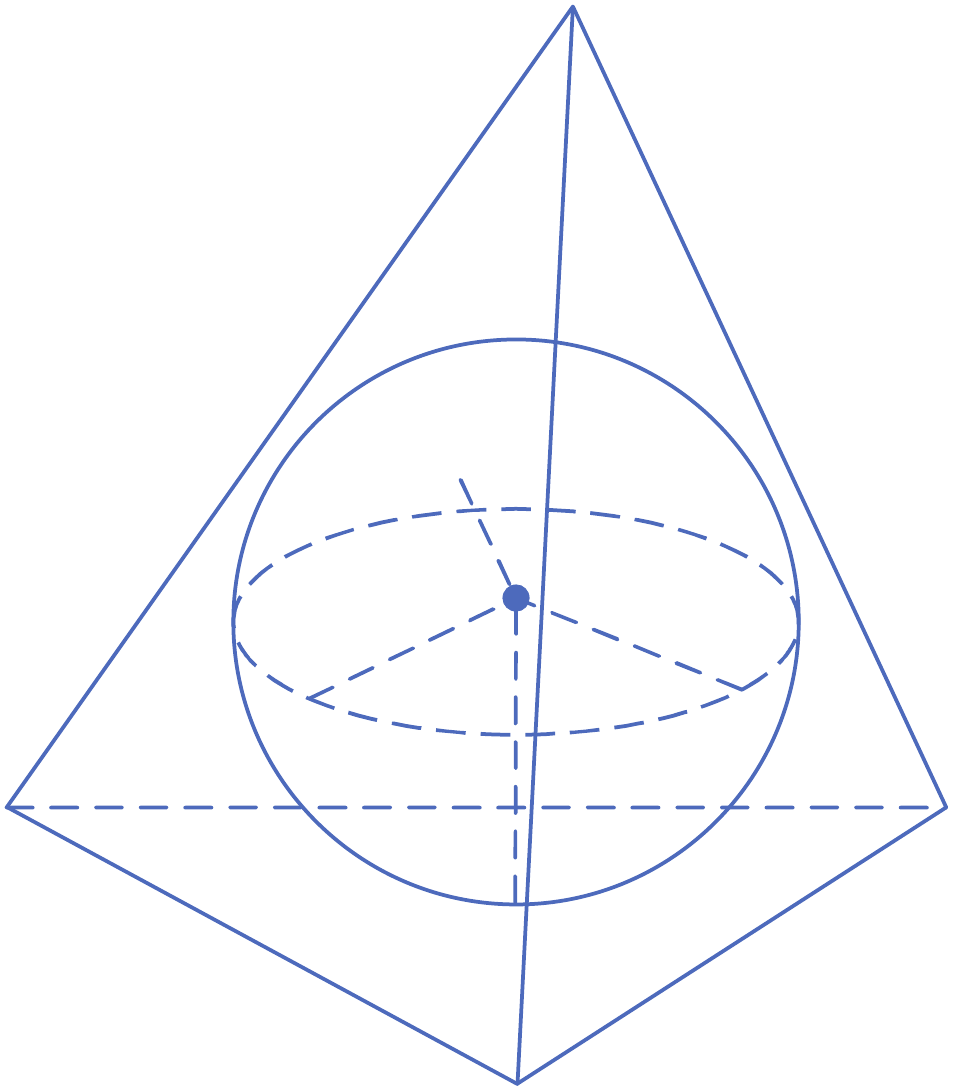}\\
	\caption{The geometric center of a tetrahedron}
	\label{fig_geometricCentre}
\end{figure}
In this section, we integrate the idea of geometric center into AOA localization. Two geometric center based methods are proposed to estimate the position.

For $m$th sensor, according to (\ref{angle1}), a plane containing the line $u_i u_o$ and normal to the horizontal plane is given by
\begin{align}\label{plane1}
	\tan\theta_m x-y-\tan\theta_m x_m+y_m=0.
\end{align}
Similarly, referring to (\ref{angle2}), a plane containing the line $u_i u_o$ and making an angle of $\phi_m$ with the horizontal plane can be expressed as
\begin{align}\label{plane2}
	&\tan\phi_m\cos\theta_mx+\tan\phi_m\sin\theta_my-z \nonumber\\
	&~~~~~-\tan\phi_m\cos\theta_mx_m-\tan\phi_m\sin\theta_m y_m+z_m=0.
\end{align}
For an arbitrary fixed point in the space $\mathbf{\tilde{u}}=[\tilde{x},\tilde{y},\tilde{z}]^T$, the values of Euclidean distance to the above two surfaces, (\ref{plane1}) and (\ref{plane2}), are given by (\ref{distance1}) and (\ref{distance2}), which are shown at the top of this page.
\begin{figure*}[ht] 
\begin{align}\label{distance1}
	d_{m,\theta}(\mathbf{\tilde{u}})=\frac{|\tan\theta_m \tilde{x}-\tilde{y}-\tan\theta_m x_m+y_m|}{\sqrt{(\tan\theta_m)^2+1}},
\end{align}
\begin{align}\label{distance2}
	d_{m,\phi}(\mathbf{\tilde{u}})=\frac{|\tan\phi_m\cos\theta_m\tilde{x}+\tan\phi_m\sin\theta_m\tilde{y}-\tilde{z}-\tan\phi_m\cos\theta_mx_m-\tan\phi_m\sin\theta_my_m+z_m|}{ \sqrt{(\tan\phi_m\cos\theta_m)^2+(\tan\phi_m\sin\theta_m)^2+1}}
\end{align}
\vspace*{6pt}
\hrulefill
\end{figure*}
Then, to simplify the expressions of (\ref{distance1}) and (\ref{distance2}), let us define
\begin{equation}\label{a_thetai}
	\mathbf{a}_{m,\theta}= \left[\sin \theta_m,-\cos \theta_m,0\right]^T
\end{equation}
\begin{equation}\label{b_thetai}
	b_{m,\theta}= \sin \theta_m x_m - \cos \theta_m y_m
\end{equation}
\begin{align}\label{a_phii}
	\mathbf{a}_{m,\phi}=
	\left[\sin\phi_m\cos\theta_m,\sin\phi_m\sin\theta_m,-\cos\phi_m \right]^T
\end{align}
\begin{equation}\label{b_phii}
	b_{m,\phi}=\sin\phi_m\cos\theta_m x_m+\sin\phi_m\sin\theta_m y_m-\cos\phi_m z_m.
\end{equation}
Thus, (\ref{distance1}) and (\ref{distance2}) can be rewritten as some direct matrix manipulations, which are given by
\begin{align}\label{distance1_c}
	d_{m,\theta}(\mathbf{\tilde{u}})=\left\|\mathbf{a}^T_{m,\theta} \mathbf{\tilde{u}}-b_{m,\theta}\right\|
\end{align}
\begin{align}\label{distance2_c}
	d_{m,\phi}(\mathbf{\tilde{u}})=\left\|\mathbf{a}^T_{m,\phi} \mathbf{\tilde{u}}-b_{m,\phi}\right\|
\end{align}

\subsection{Center of the inscribed sphere}
Let us denote the radius of inscribed sphere as $r$. According to the property of the inscribed sphere, distances from the centre to all planes are equal to $r$. In other words, $d_{m,\theta}=d_{m,\phi}=r$. To minimize the squared sum of errors, the cost function can be written as
\begin{align}\label{AOAr2P1}
	\min_{\mathbf{u},r}~~\sum_{m=1}^{M}\left(d_{m,\theta}(\mathbf{u})-r\right)^2+\left(d_{m,\phi}(\mathbf{u})-r\right)^2
\end{align}
Referring to \cite{convexInSignal}, (\ref{AOAr2P1}) is a nonsmooth, nonconvex problem. Fortunately, standard fixed-point (SFP) scheme, belonging to the class of majorization-minimization approach. Firstly, partial derivatives of (\ref{AOAr2P1}) with respect to $\mathbf{u}$ and $r$ are respectively given by
\begin{align}\label{pfpr}
	\frac{\partial f\left(\mathbf{u},r\right)}{\partial r}=&4Mr\nonumber\\
&-2\sum_{m=1}^{M}\left\|\mathbf{a}^T_{m,\theta} \mathbf{\mathbf{u}}-b_{m,\theta}\right\|+\left\|\mathbf{a}^T_{m,\phi} \mathbf{\mathbf{u}}-b_{m,\phi}\right\|
\end{align}
and
\begin{align}\label{pfpu}
	&\frac{\partial f\left(\mathbf{u},r\right)}{\partial \mathbf{u}} \nonumber\\
	&=2\sum_{m=1}^{M}\mathbf{a}_{m,\theta}\left(\mathbf{a}^T_{m,\theta}\mathbf{\mathbf{u}}-b_{m,\theta}\right)- r\mathbf{a}_{m,\theta} \frac{\mathbf{a}^T_{m,\theta}\mathbf{\mathbf{u}}-b_{m,\theta}}{\left\|\mathbf{a}^T_{m,\theta}\mathbf{\mathbf{u}}-b_{m,\theta}\right\|}\nonumber\\
	&~~~+2 \sum_{m=1}^{M}\mathbf{a}_{m,\phi}\left(\mathbf{a}^T_{m,\phi}\mathbf{\mathbf{u}}-b_{m,\phi}\right) - r\mathbf{a}_{m,\phi} \frac{\mathbf{a}^T_{m,\phi}\mathbf{\mathbf{u}}-b_{m,\phi}}{\left\|\mathbf{a}^T_{m,\phi}\mathbf{\mathbf{u}}-b_{m,\phi}\right\|}.
\end{align}
Then, according to SFP, the solution of (\ref{AOAr2P1}) can be obtained by following iterations
\begin{align}\label{r_equal}
	r_{k+1}=\frac{1}{2M}\sum_{m=1}^{M}\left\|\mathbf{a}^T_{m,\theta} \mathbf{u}_k-b_{m,\theta}\right\|+\left\|\mathbf{a}^T_{m,\phi} \mathbf{u}_k-b_{m,\phi}\right\|
\end{align}
\begin{align}\label{u_k}
	\mathbf{u}_{k+1} = \tilde{\mathbf{A}}^{-1}\left( \tilde{\mathbf{b}}+r_{k+1}\tilde{\mathbf{a}}(\mathbf{u}_k)\right)
\end{align}
where
\begin{align}\label{A_tilde}
    \tilde{\mathbf{A}} = \sum_{m=1}^{M} \mathbf{a}_{m,\theta}\mathbf{a}_{m,\theta}^T+\mathbf{a}_{m,\phi}\mathbf{a}_{m,\phi}^T,
\end{align}
\begin{align}\label{b_tilde}
    \tilde{\mathbf{b}} = \sum_{m=1}^{M} \mathbf{a}_{m,\theta}b_{m,\theta}+\mathbf{a}_{m,\phi}b_{m,\phi}
\end{align}
and
\begin{align}\label{a_tilde}
    \tilde{\mathbf{a}}(\mathbf{u}) = \sum_{m=1}^{M}\mathbf{a}_{m,\theta} \frac{\mathbf{a}^T_{m,\theta}\mathbf{\mathbf{u}}-b_{m,\theta}}{\left\|\mathbf{a}^T_{m,\theta}\mathbf{\mathbf{u}}-b_{m,\theta}\right\|} +\mathbf{a}_{m,\phi} \frac{\mathbf{a}^T_{m,\phi}\mathbf{\mathbf{u}}-b_{m,\phi}}{\left\|\mathbf{a}^T_{m,\phi}\mathbf{\mathbf{u}}-b_{m,\phi}\right\|}.
\end{align}
So far, $\mathbf{u}_0$ and $r$ can be estimated by iteration. 

Let us now analyze the radius of inscribed sphere. Given the angle error is small, we have the approximation
\begin{equation}\label{angle_appro1}
	\sin(\theta_m+n_{\theta_m})\approx \sin\theta_m + n_{\theta_m}\cos\theta_m
\end{equation}
\begin{equation}\label{angle_appro2}
	\cos(\theta_m+n_{\theta_m})\approx \cos\theta_m - n_{\theta_m}\sin\theta_m
\end{equation}
$r$ could be expressed as
\begin{equation}\label{r_ext}
	r = \frac{1}{2M}\sum_{m=1}^{M}\mathbb{E}\left[\left\|\mathbf{a}^T_{m,\theta} \mathbf{u}-b_{m,\theta}\right\|+\left\|\mathbf{a}^T_{m,\phi} \mathbf{u}-b_{m,\phi}\right\| \right]
\end{equation}
Thus, substituting (\ref{angle_appro1}) and (\ref{angle_appro2}) into (\ref{r_ext}), we have
\begin{align}\label{r_sub1}
	\left\|\mathbf{a}^T_{m,\theta+n_{\theta_m}} \mathbf{u}-b_{m,\theta}\right\| &= \left\|n_{\theta_m}((x_0 - x_m)\cos\theta_m\right. \nonumber \\
	&~~~\left.+(y_0 - y_m)\sin\theta_m)\right\| \nonumber \\
	&= \left\|n_{\theta_m}d_m\cos\phi_m\right\|
\end{align}
\begin{align}\label{r_sub2}
	&\left\|\mathbf{a}^T_{m,\phi} \mathbf{u}-b_{m,\phi}\right\| \nonumber \\
	&~~~~= \left\| \sin\phi_m\left( (x_0-x_m)\cos\theta_m + (y_0-y_m)\sin\theta_m \right) \right.\nonumber \\
	&~~~~~~~\left.- \cos\phi_m(z_0-z_m) \right\| \nonumber \\
	& ~~~~= \left\|n_{\phi_m}d_m\cos\phi_m\right\|
\end{align}
where $d_m$ is the Eucliden distance between $u_0$ and $u_m$.
Thus, $r$ can be reduced to a simple form as
\begin{equation}\label{r_equ}
	r = \frac{1}{2M}\sum_{m=1}^M{\sigma_\theta d_m \|\cos\phi_m\| + \sigma_\phi d_m}
\end{equation}
It is obvious that the value of $r$ is related to the measurement noise.
Consider a special case, $\sigma^2_{\theta}=\sigma^2_{\phi}=\sigma^2$, the $\sigma^2$ can be estimated by
\begin{align}\label{sigma2}
	\hat{\sigma^2} = \frac{4M^2r^2}{\sum_{m=1}^M{d_m^2\left( \cos^2\phi_m + 1\right)}}
\end{align}
Hence, this method can estimate the source location and variance of angle measurement error at the same time. We shall call this method as the center of the inscribed sphere (CIS) method.

\subsection{Minimum squared distance}
In many applications, it is not necessary to estimate the noise level. Thus, $r$ is useless in these applications.
We can simplify the (\ref{AOAr2P1}) by removing $r$, which yeilds
\begin{align}\label{AOAl1P1}
	\min_{\mathbf{u}}~~\sum_{m=1}^{M}d_{m,\theta}^2(\mathbf{u})+d_{m,\phi}^2(\mathbf{u})
\end{align}
Let us reduce (\ref{AOAl1P1}) to a simple form as
\begin{align}\label{AOAl1P2}
	\min_{\mathbf{u}}~~\|\mathbf{Au-b}\|^2
\end{align}
where
\begin{align}\label{A}
	\mathbf{A}=\left[\mathbf{A}_{\bm{\theta}i}^T,\mathbf{A}_{\bm{\phi}i}^T\right]^T
\end{align}
\begin{align}\label{b}
	\mathbf{b}=\left[\mathbf{b}_{\bm{\theta}i}^T,\mathbf{b}_{\bm{\phi}i}^T\right]^T
\end{align}
where
\begin{align}
	& \mathbf{A}_{\bm{\theta}i}=\left[\mathbf{a}_{1,\theta}, \mathbf{a}_{2,\theta}, \cdots, \mathbf{a}_{M,\theta}\right]^T \\
	& \mathbf{b}_{\bm{\theta}i}= \left[b_{1,\theta},b_{2,\theta},\cdots,b_{M,\theta}\right]^T \\
	& \mathbf{A}_{\bm{\phi}i}=\left[\mathbf{a}_{1,\phi}, \mathbf{a}_{2,\phi}, \cdots, \mathbf{a}_{m,\phi}\right]^T  \\
	& \mathbf{b}_{\bm{\phi}i}= \left[b_{1,\phi},b_{2,\phi},\cdots,b_{M,\phi}\right]^T
\end{align}
Obviously, (\ref{AOAl1P2}) is a LS problem. Hence, the solution can be given by
\begin{align}\label{u_est}
	\hat{\mathbf{u}}=\left(\mathbf{A}^T\mathbf{A}\right)^{-1}\mathbf{A}^T\mathbf{b}
\end{align}
This approach can be called as the minimum squared distance based least square (MSD-LS). Compared with CIS, This method has lower computational complexity by directly employing LS.
However, $\mathbf{A}$ in (\ref{A}) is not a noiseless matrix. In other words, both $\mathbf{A}$ and $\mathbf{b}$ contain error. Thus, to improve the accuracy, total least square (TLS) can be adopted.
%

\section{Analysis}
In this section, we show the CRLB of the 3D-AOA and compare the computational complexity of proposed methods with other related algorithms in literature.

\subsection{Performance Accuracy}
CRLB is considered as the minimum error variance of a unbiased estimator.
With the help of \cite{wang2015twc}, when the covariance matrices of angle noise are given by $\sigma^2_{\theta}\mathbf{I}_M$ and $\sigma^2_{\phi}\mathbf{I}_M$, corresponding CRLB of the 3D-AOA is given by (\ref{CRLB}). In section V, we will present the performance of proposed methods and adopt (\ref{CRLB}) as a benchmark.
\begin{figure*}[ht] 
\begin{align}\label{CRLB}
	&\mathbf{CRLB}=\nonumber\\
&\left[
\begin{array}{ccc}
  \sum\limits_{m=1}^M{\sigma^2_{\theta}\frac{\sin^2\theta_m}{d_m^2\cos^2\phi_m}}+\sigma^2_{\phi}\frac{\sin^2\phi_m\cos^2\theta_m}{d_m^2} & \sum\limits_{m=1}^M{\sigma^2_{\theta}\frac{\sin^2\phi_m\sin\theta_m \cos\theta_m}{d_m^2}-\sigma^2_{\phi}\frac{\sin\theta_m \cos\theta_m}{d_m^2\cos^2\phi_m^2}} &  -\sigma^2_{\phi}\sum\limits_{m=1}^M{\frac{\sin\phi_m \cos\phi_m\cos\theta_m}{d_m^2\cos^2\phi_m^2}}\\
  \sum\limits_{m=1}^M{\sigma^2_{\theta}\frac{\sin^2\phi_m\sin\theta_m \cos\theta_m}{d_m^2}-\sigma^2_{\phi}\frac{\sin\theta_m \cos\theta_m}{d_m^2\cos^2\phi_m^2}} & \sum\limits_{m=1}^M{\sigma^2_{\theta}\frac{\cos^2\theta_m}{d_m^2\cos^2\phi_m}+\sigma^2_{\phi}\frac{\sin^2\phi_m\sin^2\theta_m}{d_m^2}} & -\sigma^2_{\phi}\sum\limits_{m=1}^M{\frac{\sin\phi_m \cos\phi_m\sin\theta_m}{d_m^2\cos^2\phi_m^2}} \\
   -\sigma^2_{\phi}\sum\limits_{m=1}^M{\frac{\sin\phi_m \cos\phi_m\cos\theta_m}{d_m^2\cos^2\phi_m^2}} & -\sigma^2_{\phi}\sum\limits_{m=1}^M{\frac{\sin\phi_m \cos\phi_m\sin\theta_m}{d_m^2\cos^2\phi_m^2}} & \sigma^2_{\phi}\sum\limits_{m=1}^M{\frac{\cos^2\phi_m}{d_m^2}}
\end{array}
\right]^{-1}
\end{align}
\hrulefill
\end{figure*}

\subsection{Computational Complexity}

\begin{table}
	\footnotesize
	\centering
	\caption{Computational Complexity of Different Methods}
	\label{tab1}
	\scalebox{1.1}{
		\begin{tabular}{c|c}
			\hline
			\hline
			Algorithms     	& Complexity          \\
			\hline
			Conventional LS & $\mathcal{O}(23M+27)$     \\
			\hline
			MLE in \cite{wang2018twc}			& \makecell[c]{$\mathcal{O}(K(27M^3+9M^2+124M+9)$ \\ $+5M)$ }    \\
			\hline
			BR-PLE in \cite{wang2015twc}			& $\mathcal{O}(165M+897)$   \\
			\hline
			Proposed CIS 	& $\mathcal{O}(31M+7+6K(M+2))$     \\
			\hline
			Proposed MSD-LS & $\mathcal{O}(28M+27)$     \\
			\hline
			\hline
	\end{tabular}}
\end{table}

Note that values of (\ref{A_tilde}) and (\ref{b_tilde}) could be compute once in CIS. Thus, the computational complexity of CIS is $\mathcal{O}(31M+7+6K(M+2))$, where $K$ is the number of iterations. By employing LS, MSD-LS is also a low complexity method and its computational complexity is $\mathcal{O}(28M+27)$. For the ease of comparison between different methods, the computational complexity of the proposed two methods are shown in TABLE \ref{tab1}.  Meanwhile, two commonly used algorithms, LS and MLE, are also listed for comparison. Although MLE could approach the CRLB, its complexity grows at $\mathcal{O}(M^3)$. The computational complexity of the proposed two algorithms is a linear function of $M$, which is similar to LS.

\begin{figure}
	\centering
	\includegraphics[width=0.48\textwidth]{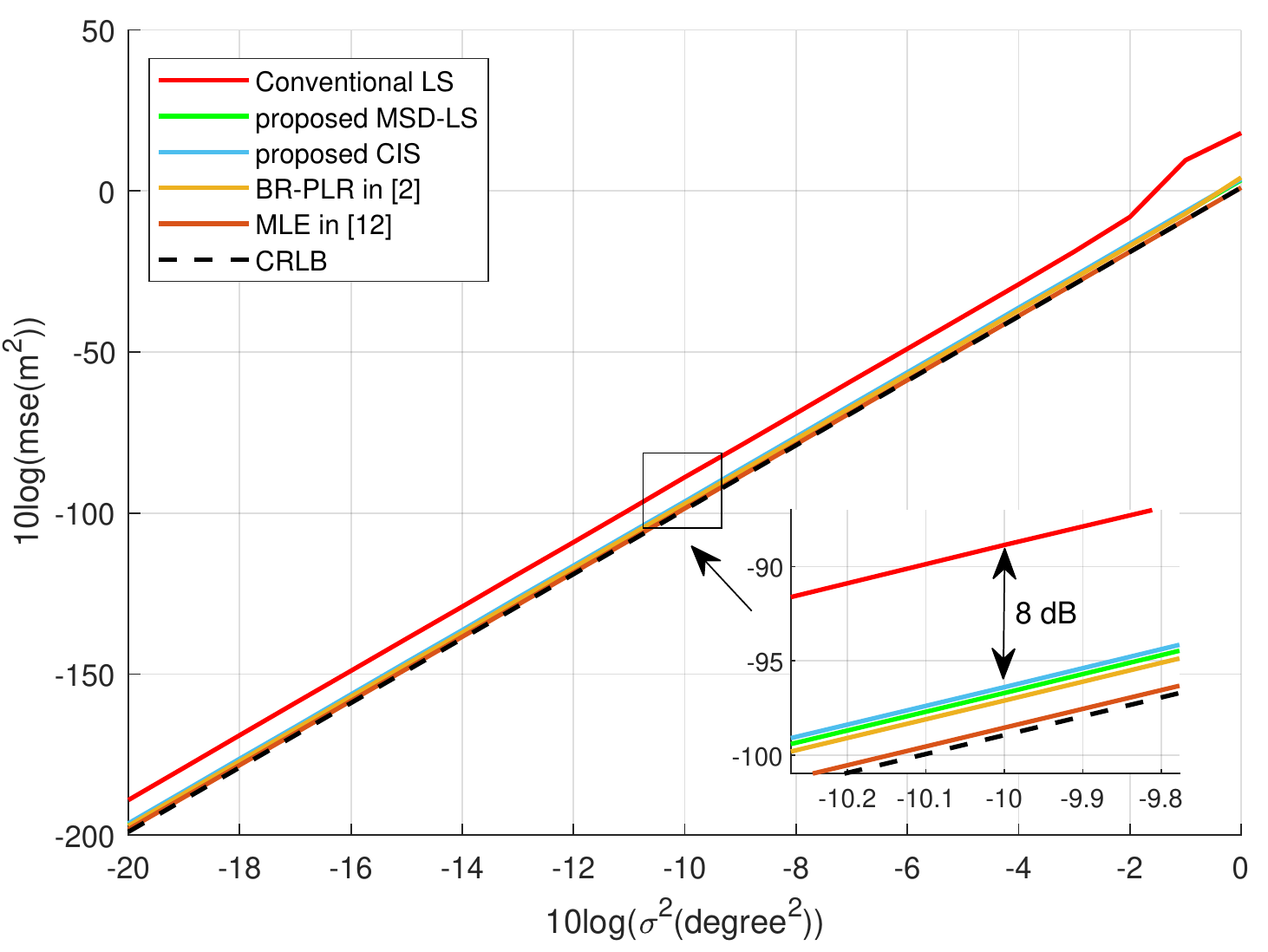}\\
	\caption{MSE of different AOA localization methods with different angle noise}
	\label{fig_diffMethods}
\end{figure}

\begin{figure}
	\centering
	\includegraphics[width=0.48\textwidth]{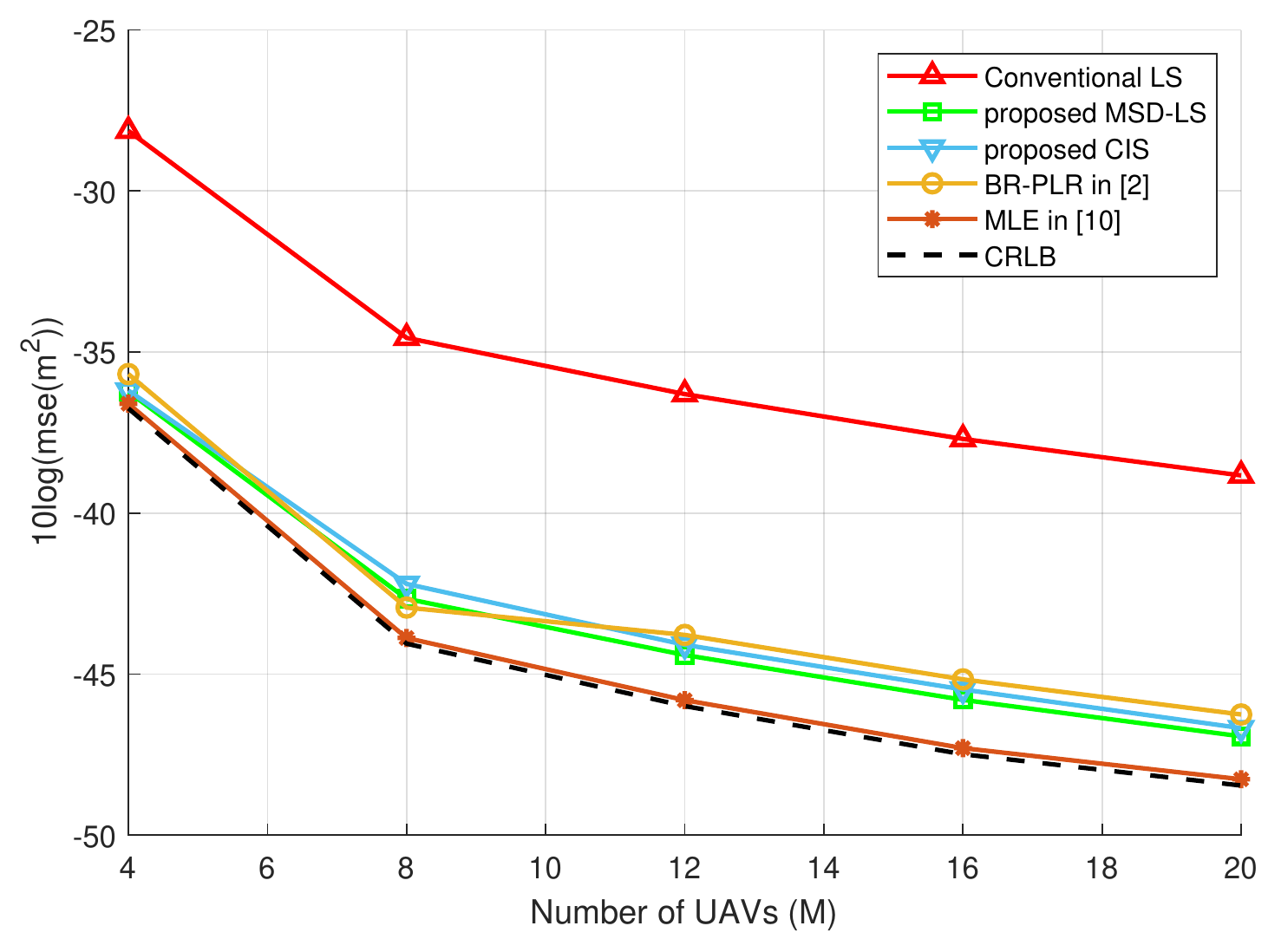}
	\caption{MSE of different AOA localization methods with different number of UAVs.\label{fig_diffM}}
\end{figure}  
\begin{figure}
	\centering
	\subfigure[Initial value is randomly generated in cube space] {
		\includegraphics[width=0.48\textwidth]{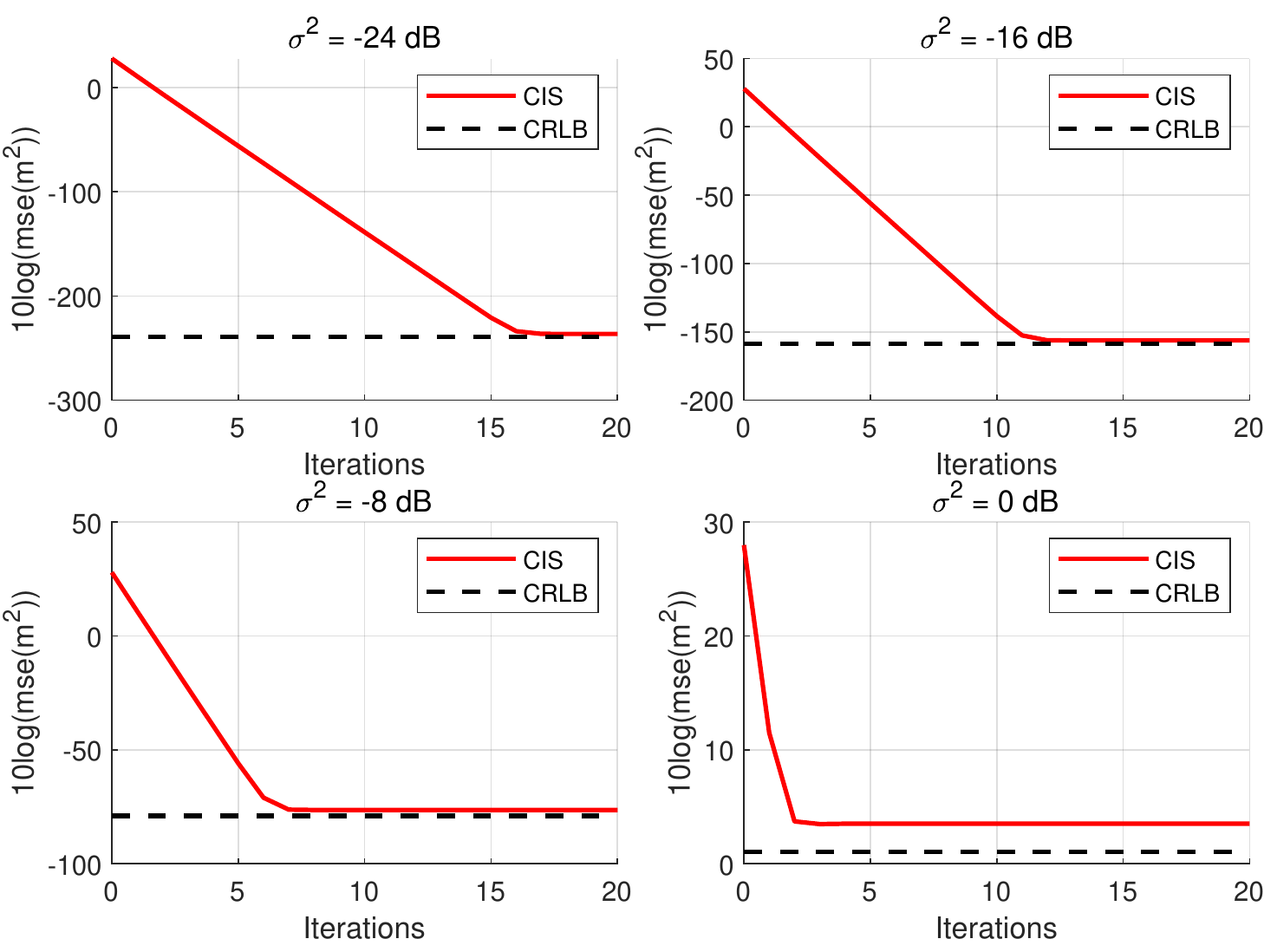}
	}\\
	\subfigure[LS is adopted is randomly generated by LS] {
		\includegraphics[width=0.48\textwidth]{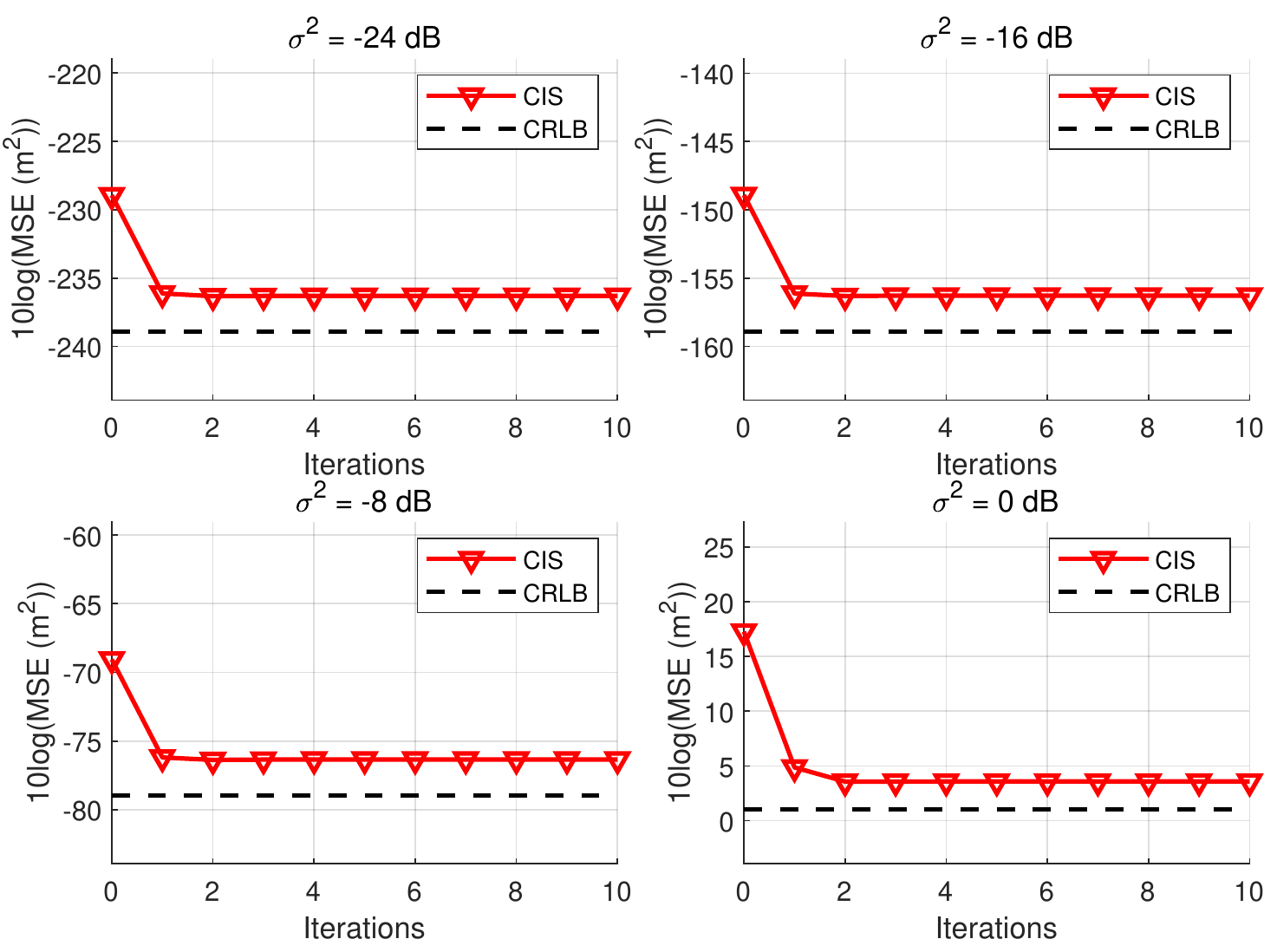}
	}
	\caption{ MSE of CIS method with iterations and different angle noise  }
	\label{fig_itNum}
	
\end{figure}
\section{Simulation Results}
In this section, simulation results are presented to evaluate the performance of different approaches. The source point is set at $[0, 0, 0]^T$. 20 sensors are randomly scattered at in a $50\times50\times50$ cube space. which is centered at $[0, 0, 0]^T$. MSE is chosen as the metric of performance, which is given by
\begin{align}\label{MSE}
	MSE=\sum_{n=1}^{N}(\hat{u}_n-u_0)^2.
\end{align}
All simulation results are averaged over 8000 Monte Carlo realizations.

In Fig.\ref{fig_diffMethods}, we demonstrate MSE of different 3D AOA methods versus the angle measurement error. All parameters and results are shown in dB. For comprasion, we consider the conventional LS in \cite{badriaslAOA2014tae}, MLE in \cite{wang2018twc} and BR-PLE in \cite{wang2015twc}. $\sigma^2$ increases from $-30$ dB to $0$ dB. MLE is almost identical with CRLB, which is accord with general knowledge. Furthermore, Proposed CIS and MSD-LS have the similar performance with BR-PLE in \cite{wang2015twc} and are able to approach the CRLB with all angle variance. In addition, compared with the LS, both of proposed methods have a gain of about 8 db. 
More importantly, referring to Section 4, proposed CIS and MSD-LS have much less computational complexity than BR-PLE and MLE.

Figure~\ref{fig_diffM} shows the MSE over number of UAVs with different 3D AOA methods. It can be seen that both of proposed methods also have a gain of about 8 db than LS and they have similar performance with BR-PLE in \cite{wang2015twc}, even better in some specail cases. This may be affected by the configuration of UAV network. Especially, when the number of UAVs becomes small, these methods could approch the CRLB. 

The MSE of CIS method versus the number of iterations with different noise is plotted in Fig.\ref{fig_itNum}, where random generator and conventional LS are adopted to generate an initial value in Fig.\ref{fig_itNum}(a) and (b), respectively. It can be seen that CIS is convergent with any initial values in Fig.\ref{fig_itNum}(a). Furthermore, convergence speed is about 16.5 dB per iteration. Thus, as noise decreases, convergence speed is faster. Observing Fig.\ref{fig_itNum}(b), we find that two or three iterations are enough when LS is used, which is faster when noise is very low. According to \ref{tab1}, when the number of iteration is over 4, it is better to choose LS as an initial solution. Therefore, random generator could be replaced by conventional LS to generate an initial value in low noise region.

\section{Conclusion}
In this work, two geometric center based low complexity approaches are proposed in 3D AOA model for UAV loclization. The location of source and angle measurement noise could be estimated by using CIS method at the same time. MSD-LS is a reduced form of CIS by deleting the angle noise estimator. Compared with the conventional LS, both of proposed methods have a gain of about 8 db, which is very close to MLE and CRLB. The computational complexity of these methods is a function of the number of UAVs, $M$, which is approach to LS and much lower than MLE. Therefore, it is very potential that our methods are widely used in 3D AOA localization for multi UAV network.
\ifCLASSOPTIONcaptionsoff
  \newpage
\fi
\bibliographystyle{IEEEtran}

\bibliography{3DAOA_ref}

\begin{thebibliography}{10}
\providecommand{\url}[1]{#1}
\csname url@samestyle\endcsname
\providecommand{\newblock}{\relax}
\providecommand{\bibinfo}[2]{#2}
\providecommand{\BIBentrySTDinterwordspacing}{\spaceskip=0pt\relax}
\providecommand{\BIBentryALTinterwordstretchfactor}{4}
\providecommand{\BIBentryALTinterwordspacing}{\spaceskip=\fontdimen2\font plus
\BIBentryALTinterwordstretchfactor\fontdimen3\font minus
  \fontdimen4\font\relax}
\providecommand{\BIBforeignlanguage}[2]{{%
\expandafter\ifx\csname l@#1\endcsname\relax
\typeout{** WARNING: IEEEtran.bst: No hyphenation pattern has been}%
\typeout{** loaded for the language `#1'. Using the pattern for}%
\typeout{** the default language instead.}%
\else
\language=\csname l@#1\endcsname
\fi
#2}}
\providecommand{\BIBdecl}{\relax}
\BIBdecl

\bibitem{shuDMRIS2021tcom}
F.~Shu, Y.~Teng, J.~Li, M.~Huang, W.~Shi, J.~Li, Y.~Wu, and J.~Wang, ``Enhanced
  secrecy rate maximization for directional modulation networks via {IRS},''
  \emph{IEEE Transactions on Communications}, vol.~69, no.~12, pp. 8388--8401,
  2021.

\bibitem{wang2015twc}
Y.~Wang and K.~C. Ho, ``An asymptotically efficient estimator in closed-form
  for {3-D} {AOA} localization using a sensor network,'' \emph{IEEE Trans.
  Wireless Commun.}, vol.~14, no.~12, pp. 6524--6535, 2015.

\bibitem{Li2021CL}
Y.~Li, F.~Shu, B.~Shi, X.~Cheng, Y.~Song, and J.~Wang, ``Enhanced {RSS}-based
  {UAV} localization via trajectory and multi-base stations,'' \emph{IEEE
  Commun. Lett.}, vol.~25, no.~6, pp. 1881--1885, 2021.

\bibitem{LiuISAC2020tcom}
F.~Liu, C.~Masouros, A.~P. Petropulu, H.~Griffiths, and L.~Hanzo, ``Joint radar
  and communication design: Applications, state-of-the-art, and the road
  ahead,'' \emph{IEEE Trans. Commun.}, vol.~68, no.~6, pp. 3834--3862, 2020.

\bibitem{shiRIS2021cl}
W.~Shi, X.~Zhou, L.~Jia, Y.~Wu, F.~Shu, and J.~Wang, ``Enhanced secure wireless
  information and power transfer via intelligent reflecting surface,''
  \emph{IEEE Communications Letters}, vol.~25, no.~4, pp. 1084--1088, 2021.

\bibitem{dongRIS2022jcn}
R.~Dong, Y.~Teng, Z.~Sun, J.~Zou, M.~Huang, J.~Li, F.~Shu, and J.~Wang,
  ``Performance analysis of wireless network aided by discrete-phase-shifter
  {IRS},'' \emph{Journal of Communications and Networks}, vol.~24, no.~5, pp.
  603--612, 2022.

\bibitem{Wufp2015tvt}
Z.-H. Wu, Y.~Han, Y.~Chen, and K.~J.~R. Liu, ``A time-reversal paradigm for
  indoor positioning system,'' \emph{IEEE Trans. Veh. Technol.}, vol.~64,
  no.~4, pp. 1331--1339, 2015.

\bibitem{cheng2022wcl}
X.~Cheng, C.~Ma, J.~Li, H.~Song, F.~Shu, and J.~Wang, ``Federated
  learning-based localization with heterogeneous fingerprint database,''
  \emph{IEEE Wireless Communications Letters}, vol.~11, no.~7, pp. 1364--1368,
  2022.

\bibitem{HoTDOA2012tsp}
K.~C. Ho, ``Bias reduction for an explicit solution of source localization
  using {TDOA},'' \emph{IEEE Trans. Signal Process.}, vol.~60, no.~5, pp.
  2101--2114, 2012.

\bibitem{Shu2018SJ}
F.~Shu, S.~Yang, J.~Lu, and J.~Li, ``On impact of earth constraint on
  {TDOA}-based localization performance in passive multisatellite localization
  systems,'' \emph{IEEE Syst. J.}, vol.~12, no.~4, pp. 3861--3864, 2018.

\bibitem{Jin2020TSP}
D.~Jin, F.~Yin, C.~Fritsche, F.~Gustafsson, and A.~M. Zoubir, ``Bayesian
  cooperative localization using received signal strength with unknown path
  loss exponent: Message passing approaches,'' \emph{IEEE Trans. Signal
  Process.}, vol.~68, pp. 1120--1135, 2020.

\bibitem{Lingren1978tae}
A.~G. Lingren and K.~F. Gong, ``Position and velocity estimation via bearing
  observations,'' \emph{IEEE Trans. Aerosp. Electron. Syst.}, vol. AES-14,
  no.~4, pp. 564--577, 1978.

\bibitem{wang2018twc}
Y.~Wang and K.~C. Ho, ``Unified near-field and far-field localization for {AOA}
  and hybrid {AOA-TDOA} positionings,'' \emph{IEEE Trans. Wireless Commun.},
  vol.~17, no.~2, pp. 1242--1254, 2018.

\bibitem{Bishop2009tae}
A.~N. Bishop, B.~D.~O. Anderson, B.~Fidan, P.~N. Pathirana, and G.~Mao,
  ``Bearing-only localization using geometrically constrained optimization,''
  \emph{IEEE Trans. Aerosp. Electron. Syst.}, vol.~45, no.~1, pp. 308--320,
  2009.

\bibitem{WeissRSS2003tvt}
A.~Weiss, ``On the accuracy of a cellular location system based on {RSS}
  measurements,'' \emph{IEEE Trans. Veh. Technol.}, vol.~52, no.~6, pp.
  1508--1518, 2003.

\bibitem{jieDOA2022wcl}
Q.~Jie, X.~Zhan, F.~Shu, Y.~Ding, B.~Shi, Y.~Li, and J.~Wang,
  ``High-performance passive eigen-model-based detectors of single emitter
  using massive {MIMO} receivers,'' \emph{IEEE Wireless Communications
  Letters}, vol.~11, no.~4, pp. 836--840, 2022.

\bibitem{shuDOA2018tcom}
F.~Shu, Y.~Qin, T.~Liu, L.~Gui, Y.~Zhang, J.~Li, and Z.~Han, ``Low-complexity
  and high-resolution {DOA} estimation for hybrid analog and digital massive
  {MIMO} receive array,'' \emph{IEEE Trans. Commun.}, vol.~66, no.~6, pp.
  2487--2501, 2018.

\bibitem{shiHADDOA2022scis}
B.~Shi, X.~Jiang, N.~Chen, Y.~Teng, J.~Lu, F.~Shu, J.~Zou, J.~Li, and J.~Wang,
  ``Fast ambiguous {DOA} elimination method of {DOA} measurement for hybrid
  massive {MIMO} receiver,'' \emph{SCIENCE CHINA Information Sciences},
  vol.~65, no.~5, pp. 159\,302--, 2022.

\bibitem{shiDOA2022sj}
B.~Shi, N.~Chen, X.~Zhu, Y.~Qian, Y.~Zhang, F.~Shu, and J.~Wang, ``Impact of
  low-resolution {ADC} on {DOA} estimation performance for massive {MIMO}
  receive array,'' \emph{IEEE Systems Journal}, vol.~16, no.~2, pp. 2635--2638,
  2022.

\bibitem{shiDOA2022wcl}
B.~Shi, L.~Zhu, W.~Cai, N.~Chen, T.~Shen, P.~Zhu, F.~Shu, and J.~Wang, ``On
  performance loss of {DOA} measurement using massive {MIMO} receiver with
  mixed-adcs,'' \emph{IEEE Wireless Communications Letters}, vol.~11, no.~8,
  pp. 1614--1618, 2022.

\bibitem{chenDOA2022wcl}
Y.~Chen, X.~Zhan, F.~Shu, Q.~Jie, X.~Cheng, Z.~Zhuang, and J.~Wang, ``Two
  low-complexity {DOA} estimators for massive/ultra-massive {MIMO} receive
  array,'' \emph{IEEE Wireless Communications Letters}, vol.~11, no.~11, pp.
  2385--2389, 2022.

\bibitem{wangAOA2012tsp}
Z.~Wang, J.-A. Luo, and X.-P. Zhang, ``A novel location-penalized maximum
  likelihood estimator for bearing-only target localization,'' \emph{IEEE
  Transactions on Signal Processing}, vol.~60, no.~12, pp. 6166--6181, 2012.

\bibitem{wang2DAOA2013wcl}
J.~Wang, J.~Chen, and D.~Cabric, ``Stansfield localization algorithm:
  Theoretical analysis and distributed implementation,'' \emph{IEEE Wireless
  Communications Letters}, vol.~2, no.~3, pp. 327--330, 2013.

\bibitem{lu2DAOA2012twc}
L.~Lu and H.-C. Wu, ``Novel robust direction-of-arrival-based source
  localization algorithm for wideband signals,'' \emph{IEEE Transactions on
  Wireless Communications}, vol.~11, no.~11, pp. 3850--3859, 2012.

\bibitem{kutluyAOA2015tsp}
K.~l. Doğancay, ``3d pseudolinear target motion analysis from angle
  measurements,'' \emph{IEEE Trans. Signal Process.}, vol.~63, no.~6, pp.
  1570--1580, 2015.

\bibitem{guiAOA2023iot}
F.~Wen, J.~Shi, G.~Gui, H.~Gacanin, and O.~A. Dobre, ``3-d positioning method
  for anonymous {UAV} based on bistatic polarized {MIMO} radar,'' \emph{IEEE
  Internet of Things Journal}, vol.~10, no.~1, pp. 815--827, 2023.

\bibitem{convexInSignal}
D.~P. Palomar and Y.~C. Eldar, \emph{Convex optimization in signal processing
  and communications}.\hskip 1em plus 0.5em minus 0.4em\relax Cambridge
  university press, 2010.

\bibitem{badriaslAOA2014tae}
L.~Badriasl and K.~Dogancay, ``Three-dimensional target motion analysis using
  azimuth/elevation angles,'' \emph{IEEE Transactions on Aerospace and
  Electronic Systems}, vol.~50, no.~4, pp. 3178--3194, 2014.

\end{thebibliography}

\end{document}